\documentclass{article}
\usepackage{latexsym}
\usepackage{graphicx}
\begin{document}
\begin{center}
\textbf{\LARGE Instantaneous Interaction between Charged Particles}
\end{center}
\begin{large}
\vspace{.3cm}
\begin{center}
W. Engelhardt\footnote{ Home address: Fasaneriestrasse 8, D-80636 
M\"{u}nchen, Germany\par Electronic address: 
wolfgangw.engelhardt@t-online.de}, retired from:
\end{center}

\begin{center}
Max-Planck-Institut f\"{u}r Plasmaphysik, D-85741 Garching, Germany
\end{center}
\vspace{1. cm} 

\noindent \textbf{Abstract} \hspace{.1cm} The interaction between charged particles through 
quasi-static fields must occur instantaneously; otherwise a violation of the 
energy principle would occur. As a consequence, the instantaneous 
transmission of both energy and information over macroscopic distances is 
feasible by using the quasi-static fields which are predicted by Maxwell's 
equations.
\vspace{1. cm} 

\noindent
\textbf{P.A.C.S.:} 41.20.Cv, 41.20.Gz, 41.20.Jb
\vspace{1. cm} 

\noindent
\textbf{Keywords:}

\begin{itemize}
\item Classical electrodynamics
\item Quasi-static electromagnetic fields
\item Transmission of information
\end{itemize}
\vspace{1. cm} 

\noindent
\textbf{I Introduction}

\noindent
Since the Special Theory of Relativity has become a constitutive part of 
physics, it is commonly believed that the transmission of energy or 
information between distant locations can only occur with the maximal speed 
of light. Maxwell's equations are usually formulated as wave equations for 
the scalar and vector potentials suggesting (in Lorenz gauge) that the 
influence of a source propagates with finite speed to a field point where an 
observer may be placed. This conception is undoubtedly true as far as 
electromagnetic waves are concerned. There exists, however, a further method 
of energy transmission effected by quasi-static fields which are also 
predicted by Maxwell's theory. In contrast to the radiation fields, they 
decay in the far field zone inversely proportional to the third power of the 
distance from the sources so that the range of transmission is much shorter 
than in case of travelling waves. As far as the magnetic quasi-static fields 
are concerned, they have a most important application for the transmission 
of energy in transformers.

Due to the linearity of Maxwell's equations the fields may be split up into 
quasi-static, instantaneous contributions, and into wave parts: $\vec 
{B}=\vec {B}_i +\vec {B}_w $, $\vec {E}=\vec {E}_C +\vec {E}_i +\vec {E}_w 
$. In case of the electric field the instantaneous part consists of the 
irrotational Coulomb field $\vec {E}_C $ and an induced rotational field 
$\vec {E}_i $. The wave fields obey inhomogeneous hyperbolic equations:
\begin{eqnarray}
\label{eq1}
\Delta \,\vec {E}_w -\frac{1}{c^2}\frac{\partial ^2\vec {E}_w }{\partial 
t^2}=\frac{1}{c^2}\frac{\partial ^2\vec {E}_i }{\partial t^2} \nonumber \\ 
 \Delta \,\vec {B}_w -\frac{1}{c^2}\frac{\partial ^2\vec {B}_w }{\partial 
t^2}=\frac{1}{c^2}\frac{\partial ^2\vec {B}_i }{\partial t^2}  
\end{eqnarray}
where we have used, e.g.: $rot\,rot\,\vec {E}_w =\nabla \left( {div\,\vec 
{E}_w } \right)-\Delta \vec {E}_w =-\Delta \vec {E}_w $. The instantaneous 
fields appear as sources in (\ref{eq1}) and can be calculated from a given charge and current distribution:
\begin{eqnarray}
\label{eq2}
\vec {E}_C \left( {\vec {x},\,t} \right)=\int\!\!\!\int\!\!\!\int \rho 
\left( {\vec {x}',\,t} \right)\,\frac{\vec {x}-\vec {x}'}{\left| {\vec 
{x}-\vec {x}'} \right|^3}\,d^3x' \nonumber \\ 
 \vec {E}_i \left( {\vec {x},\,t} \right)=-\frac{1}{4\pi 
\,c}\int\!\!\!\int\!\!\!\int {\frac{\partial \vec {B}_i \left( {\vec 
{x}',\,t} \right)}{\partial t}\times } \,\frac{\vec {x}-\vec {x}'}{\left| 
{\vec {x}-\vec {x}'} \right|^3}\,d^3x' \\ 
 \vec {B}_i \left( {\vec {x},\,t} 
\right)=\frac{1}{c}\int\!\!\!\int\!\!\!\int {\left( {\vec {j}\left( {\vec 
{x}',\,t} \right)+\frac{1}{4\pi }\frac{\partial \vec {E}_C \left( {\vec 
{x}',\,t} \right)}{\partial t}} \right)\times } \,\frac{\vec {x}-\vec 
{x}'}{\left| {\vec {x}-\vec {x}'} \right|^3}\,d^3x' \nonumber  
\end{eqnarray}
The set of equations (1, 2) is entirely equivalent to Maxwell's equations 
which may be checked by insertion into the first order system [1]. In the 
present formulation it becomes obvious that Maxwell's theory predicts not 
only wave propagation at the velocity of light (\ref{eq1}), but also instantaneous 
fields as described by the quasi-static integrals (\ref{eq2}). A complete 
cancellation of the two kinds of fields is not possible, as was pointed out 
in [2] and is obvious from (\ref{eq1}): If the sum $\vec {E}_w +\vec {E}_i $, e.g., 
would vanish for a certain interval of time, one would have $\Delta \vec 
{E}_w =0$ so that the wave field $\vec {E}_w $ would vanish everywhere when 
the usual boundary condition $\vec {E}_w \left( \infty \right)=0$ is 
imposed.

In this note we show that the transmission of energy by quasi-static fields 
must occur instantaneously as suggested by (\ref{eq2}); otherwise the energy 
principle would be violated. In Sect. II we discuss the interaction of two 
charges, and in Sect. III the interaction of two current loops. In both 
cases we come to the conclusion that energy can be transmitted 
instantaneously. 
\vspace{1. cm} 

\noindent
\textbf{II Interaction of two charges coupled by the Coulomb field}

\noindent Let us consider two positive charges which are placed at a distance $R$. In Fig. 1 charge B is rigidly attached to a heavy wall, whereas charge A can be moved by a mechanical force along a distance $r$. When charge A is pushed 
against B, a force must be applied against the repulsive electric force, and 
a certain amount of potential energy is invested. Since charge A moves in a 
conservative potential which does not vary in time, the invested work is 
recovered when the charge returns to its initial position. 

\begin{figure}[htbp]
\centerline{\includegraphics[width=5.02in,height=2.0in]{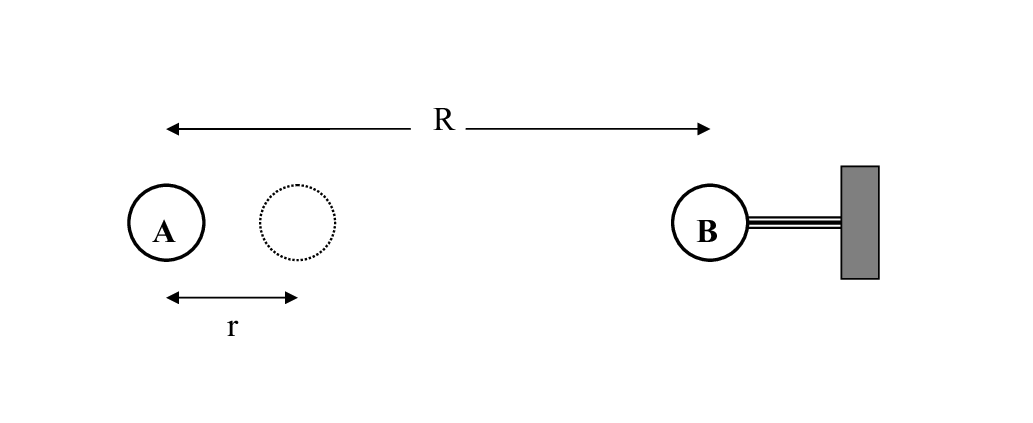}}
\caption{Interaction of two charges (charge B rigidly attached)}
\label{fig1}
\end{figure}

Next we consider Fig. 2 where charge B is attached to a flexible spring. 
Initially the spring is somewhat compressed due to the repulsive force 
between the charges. When charge A is pushed against B, the spring is 
compressed even more, and, because of its inertia, charge B will oscillate 
after charge A has returned to its initial position. 

\begin{figure}[htbp]
\centerline{\includegraphics[width=4.89in,height=1.65in]{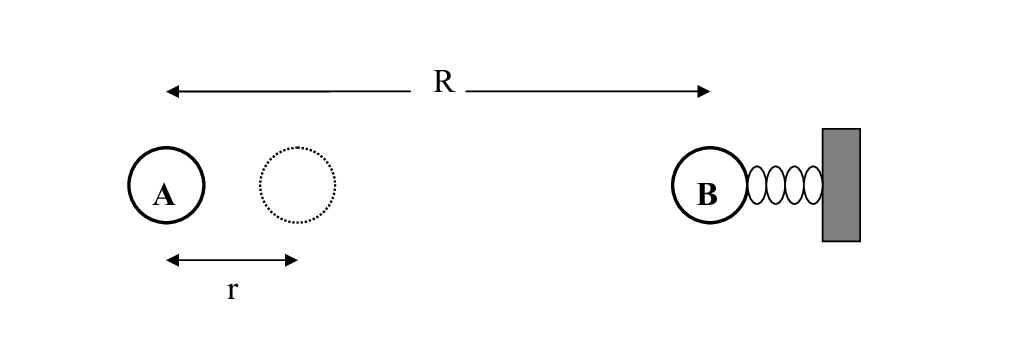}}
\caption{Interaction of two charges (charge B elastically attached)}
\label{fig2}
\end{figure}

Obviously, energy has been transmitted from A to B through the quasi-static 
electric field. The oscillation energy must have been supplied by the 
mechanical force acting on charge A. This is easily explained: Charge 
A was moving in a conservative potential which was, however, not constant in 
time, since charge B was yielding to the increased force due to its elastic 
fixture. If one carries out the exact calculation assuming that the charges 
were coupled by the Coulomb force:
\begin{equation}
\label{eq3}
F\left( t \right)=\frac{q_A {\kern 1pt}q_B }{4\pi \,\varepsilon _0 R^2\left( 
t \right)}
\end{equation}
one finds, of course, that the energy appearing in the oscillation of charge 
B was exactly supplied by the mechanical force which moved charge A.

In this seemingly trivial example we have assumed in (\ref{eq3}) that both charges 
are coupled instantaneously. Let us now assume that a time $R 
\mathord{\left/ {\vphantom {R c}} \right. \kern-\nulldelimiterspace} c$ 
elapses before charge B can ``realize'' that charge A is moving. If the 
cyclic motion of charge A is completed within a short time $\tau <R 
\mathord{\left/ {\vphantom {R c}} \right. \kern-\nulldelimiterspace} c$, 
charge B cannot react during the cycle so that charge A still moves in a 
constant conservative potential like in Fig. 1. The mechanical force does 
not produce any total work during the cycle, but after a delay time $R 
\mathord{\left/ {\vphantom {R c}} \right. \kern-\nulldelimiterspace} c$ 
charge B will start to oscillate. Its energy comes out of nothing as a 
consequence of our assumption that the action of charge A on B is delayed. 
As long as we believe in the conservation of energy we must conclude that 
the coupling is determined by (\ref{eq3}), and the energy transfer occurred 
instantaneously. 

One might argue that the acceleration of charge A produced a wave containing 
energy which was transported to B at finite velocity. It is, of course, true 
that the acceleration of charges produces waves according to Maxwell's 
theory as described by (\ref{eq1}). This holds, however, in both cases of Fig. 1 and Fig. 2. It is entirely independent of charge B being rigidly or elastically attached to the wall. In both cases the mechanical force must supply a small amount of energy which is carried away by the wave and must be accounted for in the energy balance. The salient point is, however, that the energy balance completed at charge A during a cycle cannot depend on charge B oscillating in the future or not. There must be an instantaneous feedback from charge B to charge A in order to balance correctly the work at charge A with any oscillation energy produced at charge B. This necessary feedback is provided by the Coulomb force (3), or equation (2), in general. 

Furthermore, it should be noted that the wave travels perpendicular to the acceleration of charge A and does not reach charge B at all, as it is placed in line with the acceleration vector. Thus, the production of waves cannot explain the missing energy source in case of Fig. 2, when delayed action is postulated. 

\vspace{1. cm} 

\noindent \textbf{III Interaction of two current loops}

\noindent
All the power produced by the electric companies and transmitted to the 
consumers passes several times through transformers. In this Section we show 
that the transmission from the primary to the secondary circuits must occur 
instantaneously as described by Maxwell's equations.

\begin{figure}[htbp]
\centerline{\includegraphics[width=4.60in,height=1.90in]{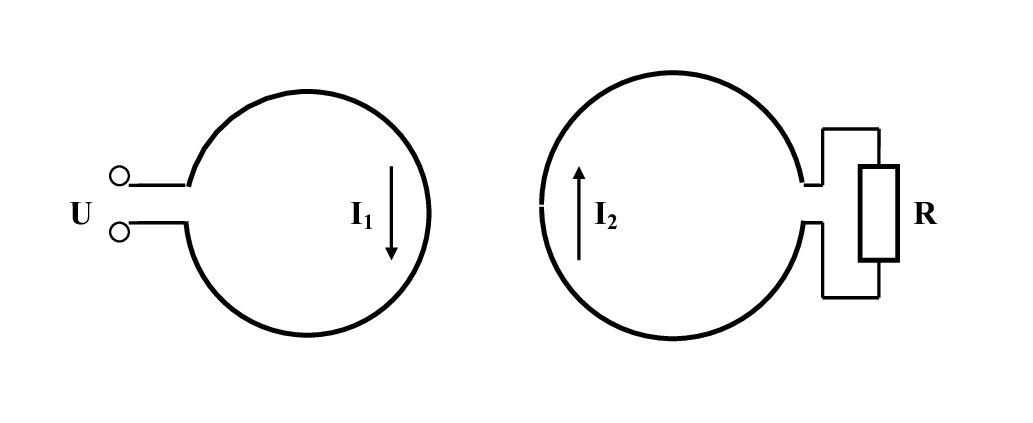}}
\caption{Interaction of two current loops}
\label{fig3}
\end{figure}

In principle, a transformer consists of two current loops as sketched in 
Fig. 3. Applying Faraday's law of induction and the laws of Amp\`{e}re and 
Ohm one has the transformer equations [3]:
\begin{equation}
\label{eq4}
U=L_1 \frac{dI_1 }{dt}+M\frac{dI_2 }{dt}
\end{equation}
\begin{equation}
\label{eq5}
0=L_2 \frac{dI_2 }{dt}+M\frac{dI_1 }{dt}+R\,I_2 
\end{equation}
where $L_{1,{\kern 1pt}2} $ are the self-inductances of the loops and $M$ is 
the coupling inductance. When there is no resistive load in the secondary 
circuit ($R=0)$, it follows from (\ref{eq5}): $L_2 \,I_2 +M\,I_1 =0$ for alternating 
currents. $I_1 $ and $I_2 $ are out of phase by 180 degrees. From (\ref{eq4}) 
follows then that there is a phase difference of 90 degrees between the 
applied voltage $U$ and the current $I_1 $. The power $U\,I_1 $ injected 
into the primary circuit oscillates forth and back so that no energy is 
deposited into the (ideal) transformer on average. If the secondary 
resistance is finite, a phase shift occurs which may be calculated by 
solving the differential equations (\ref{eq4}) and (\ref{eq5}). As a result the time 
integral $\int {U\,I_1 {\kern 1pt}dt} $ does not vanish anymore, but 
supplies the energy $\int {R\,I_2^2 \,dt} $ which is dissipated in the 
secondary circuit. Energy is obviously transmitted over the distance between 
the loops. Since in (4, 5) Maxwell's displacement current was neglected, the 
coupling of the loops was assumed to be instantaneous which holds then also 
for the energy transfer. 

If we would assume that it takes some time for the magnetic field produced 
in loop 1 to travel to loop 2, induce a current there which in turn produces 
a magnetic field travelling back to loop 1, one would have a phase shift of 
more than 90 degrees between voltage and current in the primary circuit, 
even in the case of vanishing resistance ($R=0)$. The integral $\int {U\,I_1 
{\kern 1pt}dt} $ would not be zero on average, and energy would be lost in 
the ideal transformer which is supposed to contain only superconducting 
loops. As in the previous Section, we must conclude that the coupling in a 
transformer cannot be effected by an electromagnetic wave, but must be 
caused by quasi-static instantaneous fields, in order to avoid a clash with 
the energy principle. The quasi-static magnetic field can apparently be used 
like the quasi-static electric field to transmit information faster than 
light. 

In industrial transformers the distance between the primary and the secondary circuit is chosen to be very small, but one can arrange the two loops of Fig. 3 at an appreciable distance in order to measure the time of transmission. This was in fact done by Kholmetskii and coworkers [4]. They found experimentally that a ``bound'' magnetic field as described by (2) is spreading at a velocity ``highly exceeding the velocity of light''.

\vspace{1. cm} 

\noindent
\textbf{IV Conclusion}

\noindent
In two simple examples it was demonstrated that the coupling of electric 
charges and currents through quasi-static fields must occur instantaneously, 
in order to maintain the conservation of energy. In technical applications 
this kind of coupling is assumed anyway, but it is frequently thought that 
engineers just use a practical approximation, whereas the `correct' interaction requires a description in terms of travelling wave fields. Our analysis 
shows that this is not the case. It proves that instantaneous transmission 
of information over macroscopic distances is possible in agreement with 
Maxwell's theory and with very recent experiments. The argument brought forward in Sect. II may also be applied to the mechanical force acting between the current loops of Fig. 3. It could be extended to the gravitational force as well.

\end{large}

\vspace{1. cm} 

\newpage

\begin{thebibliography}{99}
\bibitem{ Jackson }
{ J. D. Jackson, \textit{Classical Electrodynamics, }Second Edition, John Wiley {\&} Sons, Inc., New York (1975).}
\bibitem{ Engelhardt } { W. Engelhardt, \textit{Gauge invariance in classical electrodynamics, } Annales de la Fondation Louis de Broglie, Vol. 30 No. 2, page 157 (2005). }
\bibitem{ Becker } { R. Becker, F. Sauter, \textit{Theorie der Elektrizit\"{a}t, }Erster Band, B. G. Teubner Verlagsgesellschaft, Stuttgart (1962). }
\bibitem{ Kholmetskii } {A.L. Kholmetskii, O.V. Missevitch, R. Smirnov-Rueda, R.I. Tzontchev, A.E. Chubykalo, I. Moreno, \textit{ Experimental Evidence on Non-Applicability of the Standard Retardation Condition to Bound Magnetic Fields and on New Generalized Biot-Savart Law }, arXiv:physics/0601084 v1, 12. Jan. 2006 }
 
\end{thebibliography}
\end{document}